\pdfoutput=1
\documentclass[usenatbib]{mn2e}
\usepackage{apjfonts}
\usepackage{amssymb}
\usepackage{amsmath}
\usepackage{ctable}
\usepackage{fixltx2e} 

\newcommand{\be}{\begin{equation}}
\newcommand{\ee}{\end{equation}}

\newcommand{\msun}{M_{\sun}}

\newcommand{\flast}{f_{\ell}}
\newcommand{\Sprime}{S^{\prime}}
\newcommand{\HC}{HC08}
\newcommand{\paperone}{H11}

\newcommand\plotonesize[2]
 {\centering \leavevmode \includegraphics[width={#2\columnwidth}]{#1}}

\newcommand{\acknowledgments}{\begin{small}\section*{Acknowledgments}\end{small}}
\newcommand\altaffilmark[1]{$^{#1}$}
\newcommand\altaffiltext[1]{$^{#1}$}
\title[The CMF \&\ Stellar IMF]{The Stellar IMF, Core Mass Function, \&\ The Last-Crossing 
Distribution\vspace{-0.5cm}}

\author[Hopkins]{
\parbox[t]{\textwidth}{ 
Philip F. Hopkins\altaffilmark{1}\thanks{E-mail:phopkins@astro.berkeley.edu}} 
\vspace*{6pt} \\
\altaffiltext{1}{Department of Astronomy, University of California
  Berkeley, Berkeley, CA 94720\vspace{-1.1cm}} \\
}

\date{Submitted to MNRAS, December, 2011\vspace{-0.6cm}}
\begin{document}
\maketitle
\label{firstpage}

\begin{abstract}

\citet{hennebelle:2008.imf.presschechter} (\HC) attempted to derive the stellar IMF as a consequence 
of lognormal density fluctuations in a turbulent medium, using an argument 
similar to \citet{pressschechter} for Gaussian random fields. 
Like that example, however, the solution there does not resolve the 
``cloud in cloud'' problem; it also does not extend to 
the large scales that dominate the velocity and density fluctuations.
In principle, these can change the results at 
the order-of-magnitude level or more. In this paper, we use the results 
from \citet{hopkins:excursion.ism} (\paperone) to generalize the excursion set 
formalism and derive the exact solution in this regime. 
We argue that the stellar IMF and core mass function (CMF) should be associated with the {\em last-crossing} 
distribution, i.e.\ the mass spectrum of bound objects defined on the {\em smallest} 
scale on which they are self-gravitating. This differs 
from the {\em first-crossing} distribution (mass function on the largest self-gravitating 
scale) which is defined in cosmological 
applications and which \paperone\ show corresponds to the GMC 
mass function in disks. We derive an analytic equation for the last-crossing 
distribution that can be applied for an arbitrary collapse threshold shape in 
ISM and cosmological studies. With this, we show that the {\em same} model 
that predicts the GMC mass function and large-scale structure of galaxy disks 
{\em also} predicts the CMF -- and by extrapolation stellar IMF -- in good agreement with observations. The only adjustable parameter in the model is the turbulent velocity power spectrum, 
which in the range $p\approx5/3-2$ gives similar results. 
We also use this to formally justify why the approximate solution 
in \HC\ is reasonable (up to a normalization constant) 
over the mass range of the CMF/IMF; however there are 
significant corrections at intermediate and high masses. 
We discuss how the exact solutions here can be used to 
predict additional quantities such as the clustering of stars, 
and embedded into time-dependent models that follow 
density fluctuations, fragmentation, mergers, and successive generations 
of star formation.

\end{abstract}

\begin{keywords}
star formation: general --- galaxies: formation --- galaxies: evolution --- galaxies: active --- 
cosmology: theory
\vspace{-1.0cm}
\end{keywords}

\vspace{-1.1cm}
\section{Introduction}
\label{sec:intro}

The origin of the stellar initial mass function (IMF) is a question of fundamental 
importance for the study of star formation, stellar evolution and feedback, 
and galaxy formation. It is an input into a huge range of models of 
all of these phenomena, and a necessary assumption when deriving 
physical parameters from many observations. However, despite decades of 
theoretical study, it remains poorly understood. A critical first step -- although by no means a complete description of the origin of the IMF -- is understanding the origin and form of the mass function of protostellar cores (the CMF), specifically that of self-gravitating, collapsing cores that will ultimately form stars.

Recently, \HC\ presented a compelling argument for the physical origin of the IMF shape, as a consequence of the CMF resulting from lognormal density fluctuations in a turbulent medium. It is increasingly clear that the density structure of the ISM is dominated by supersonic turbulence over a wide range of scales \citep[e.g.][]{scalo:2004.turb.fx.review,elmegreen:2004.obs.ism.turb.review,mac-low:2004.turb.sf.review,mckee:2007.sf.theory.review}, and a fairly generic consequence of this is that the density distribution converges towards a lognormal PDF, with a dispersion that scales weakly with Mach number \citep[e.g.][]{vazquez-semadeni:1994.turb.density.pdf,padoan:1997.density.pdf,scalo:1998.turb.density.pdf,ostriker:1999.density.pdf}. Based on the analogy between this and cosmological Gaussian density fluctuations, \HC, building on the earlier work in \citet{inutsuka:2001.mol.core.mf.eps} as well as \citet{padoan:1997.density.pdf,padoan:2002.density.pdf}, attempted to approximate the mass function of self-gravitating cores in a manner exactly analogous to \citet{pressschechter}. If the density field $\delta({\bf x})$ (where $\delta\propto \ln{\rho}$) at the point ${\bf x}$ is normally distributed, then the average $\delta$ smoothed around the point ${\bf x}$ with the appropriate window function of effective radius $R$, $\delta({\bf x}\,|\,R)$ is also normally distributed, with a variance $S=\sigma^{2}(R)$ that can (in principle) be calculated from the power spectrum or simply estimated from the Mach number \citep[see e.g.][]{passot:1998.density.pdf,nordlund:1999.density.pdf.supersonic,federrath:2008.density.pdf.vs.forcingtype,price:2011.density.mach.vs.forcing}. The total mass which is self-gravitating, over the scale $R$, is then just $\propto \int_{B(R)}^{\infty} \rho(\delta)\,P(\delta){\rm d}\delta$, where $B(R)$ is the critical density above which the region would be self-gravitating. Differentiating this total mass fraction with respect to the mass scale associated with each $R$ gives -- approximately -- the total mass in bound objects per unit mass, $M^{2}\,{\rm d}N/{\rm d}M$, hence the mass function. \HC\ showed that, for plausible $S$, and $B$ given by the Jeans condition for thermal plus turbulent velocities, this argument reproduces all of the key features of the observed CMF and stellar IMF (for extensions of this calculation allowing for different gas properties and , see \citealt{hennebelle:2009.imf.variation,hennebelle:2011.time.dept.imf.eps,chabrier:2011.dimsional.imf.turb.args}).

While extremely interesting, 
there are, however, a number of uncertain assumptions in this derivation of the CMF. Because it focuses on small scales exclusively, the properties of the ``parent'' scales (e.g.\ GMCs) must be assumed somewhat ad hoc. Since most of the power in velocity (hence density) fluctuations is on large scales (true for any reasonable turbulent power spectrum, and also observed; see \citealt{ossenkopf:2002.obs.gmc.turb.struct,brunt:2009.turb.driving.gmc.obs}), this means $S(R)$ could also not be calculated but was instead assumed, and its ``run'' with radius $R$ (being undetermined) was neglected (and the resulting mass function artificially truncated before going to very large scales). 

Most important, like \citet{pressschechter}, this derivation does not resolve the ``cloud in cloud'' problem. 
A given region may well be self-gravitating on many different scales $R$ (the smoothed $\delta({\bf x}\,|\,R)$ 
crossing back and forth across the critical $B(R)$ with scale). This makes the resulting mass function inherently ambiguous, since different crossings of the same spatial location are counted multiple times and with different signs. \citet{bond:1991.eps} resolved this ambiguity by extending the mathematical excursion set formalism and defining the ``first crossing distribution.'' This allowed them to rigorously calculate the mass function of objects -- counting each point only once -- where ``mass'' was defined as the mass enclosed in the {\em largest} scale $R$ on which a region was self-gravitating. 

More recently, \paperone\ showed how the full excursion set formalism could be generalized to the problem of lognormal density fluctuations in a gaseous galactic disk. By including disk scale-effects, they showed that it is also possible to predict the absolute mass scales, variance $\sigma^{2}(R)$, and barrier $B(R)$ -- quantities which had to be assumed {\em ad hoc} in \HC\ --  with only the assumption of a turbulent spectral shape. On small scales (below the disk scale height), the collapse condition is just the Jeans condition for turbulent plus thermal support -- identical to \HC -- so there is no reason why the approach therein cannot be extended to the same scales. 

\paperone\ showed that the first-crossing distribution for turbulent gas in a galaxy disk is {\em not} the CMF or stellar IMF, but rather agrees extremely well with observations of the mass function and other properties of giant molecular clouds (GMCs). This highlights just how important the distinction of multiple crossings can be -- extending the \HC\ argument to large scales, self-gravitating objects with scales $\sim10^{6}\,\msun$ would be recovered (and contain more mass than the objects self-gravitating on scales $\sim 1\,\msun$). Clearly, these are the ``parent'' clouds, not the protostellar cores! 

So how can the two be distinguished? Physically, consider a region which is self-gravitating on a large scale $R_{0}$. If it contains multiple sub-regions that are themselves self-gravitating on a smaller scale $R_{1}$, then the entire $R_{0}$ object will not form a single core \citep[see e.g.\ the discussion of the ``cloud in cloud'' problem in][]{veltchev:2011.frag,donkov:2011.frag}. Since the mean density at $R_{1}$ to be self-gravitating must be larger than that at $R_{0}$, these sub-regions will collapse more rapidly -- the ``parent cloud'' is fragmenting into smaller objects. This can be continued iteratively inside $R_{1}$. It is only when a region is self-gravitating on a scale $R_{\ell}$, and {\em not} self-gravitating on any smaller scales, that its collapse will proceed without fragmentation.

Therefore, we argue in this paper that the CMF (and, to the extent that it is related, the IMF) should be associated with the {\em last-downcrossing} distribution: i.e.\ the mass function of regions which are self-gravitating, but with mass defined at the smallest scales on which they are self-gravitating. 
This has not, to our knowledge, been studied in any cosmological context (since halos are assumed to collapse ``into'' the mass they contain), so in \S~\ref{sec:lastcross}, we derive a rigorous analytic expression for this mass function as a function of arbitrary collapse threshold. 
In \S~\ref{sec:solutions} we combine this with the model from \paperone, which gives the appropriate collapse threshold and variance for a galactic disk.
In \S~\ref{sec:results} we compare the results to the observed CMF and stellar IMF, see how it contrasts with the first-crossing distribution (GMC mass function), and examine its dependence on turbulent properties. 
In \S~\ref{sec:discussion} we discuss implications and future directions for this work.

\begin{figure}
    \centering
    \plotonesize{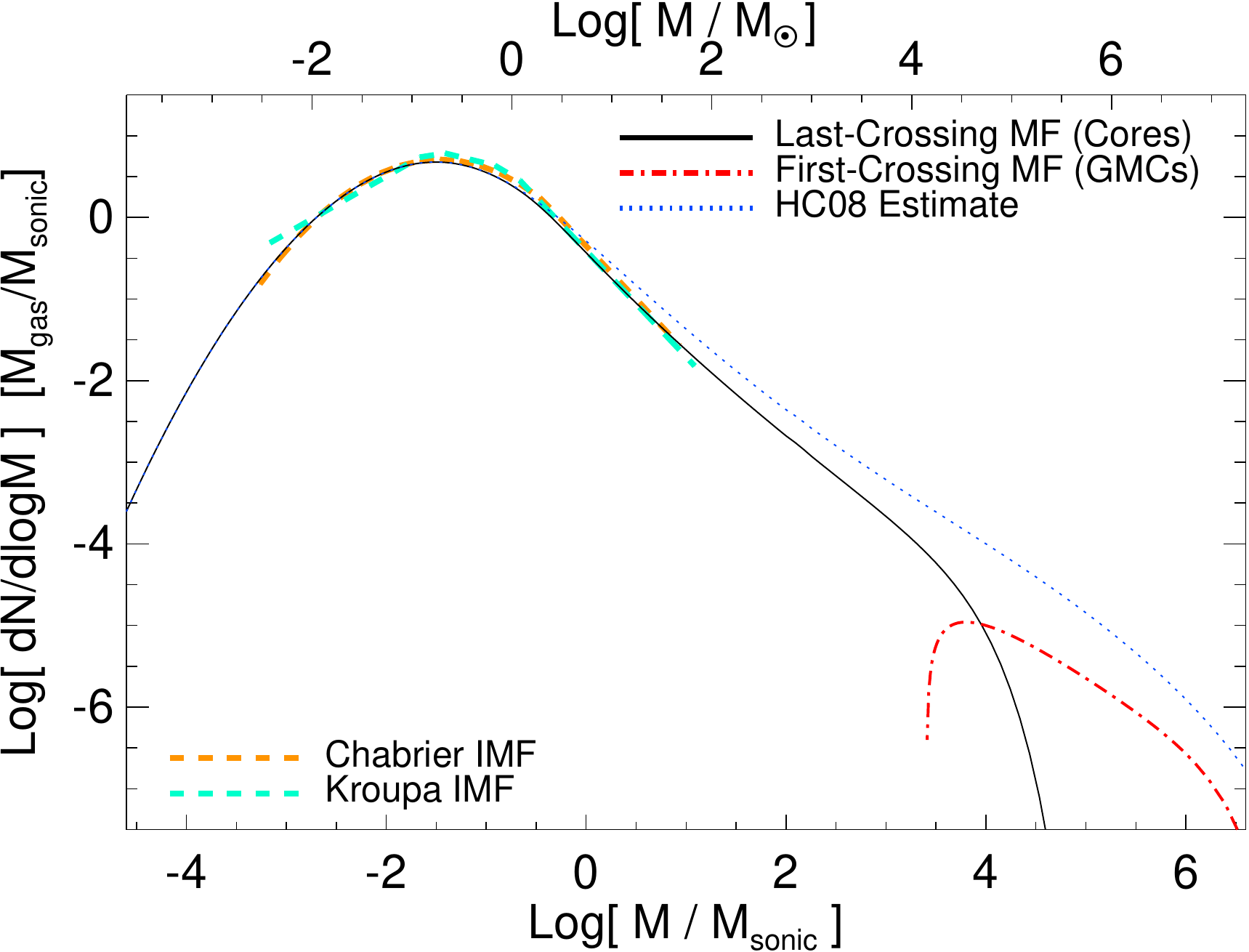}{0.95}
    \caption{Predicted last-crossing mass function 
    (distribution of bound masses measured on the 
    {\em smallest} scale on which they are self-gravitating) 
    for a galactic disk with turbulent spectral index $p=2$, 
    and Mach number at scale $\sim h$ of $\mathcal{M}_{h}=30$. In units of 
    the sonic mass $M_{\rm sonic}\equiv (2/3)\,c_{s}^{2}\,R_{\rm sonic}/G$ 
    and total disk gas mass $M_{\rm gas}$, all other 
    properties are completely specified by disk stability. 
    We calculate this with the analytic iterative solution 
    to Eq.~\ref{eqn:flast}; the standard Monte Carlo excursion set method 
    gives an identical result.
    We compare the first-crossing distribution -- the distribution of 
    masses measured on the {\em largest} scale on which 
    gas is self-gravitating, which in \paperone\ we 
    show agrees extremely well with observed GMC mass functions.
    The MF derived in \HC\ by ignoring multiple crossings is 
    also shown. We compare the observed stellar IMF from \citet{kroupa:imf} and \citet{chabrier:imf} (shifted to higher masses by a simple core-to-stellar mass factor of $3$). 
    \label{fig:imf}}
\end{figure}

\begin{figure}
    \centering
    \plotonesize{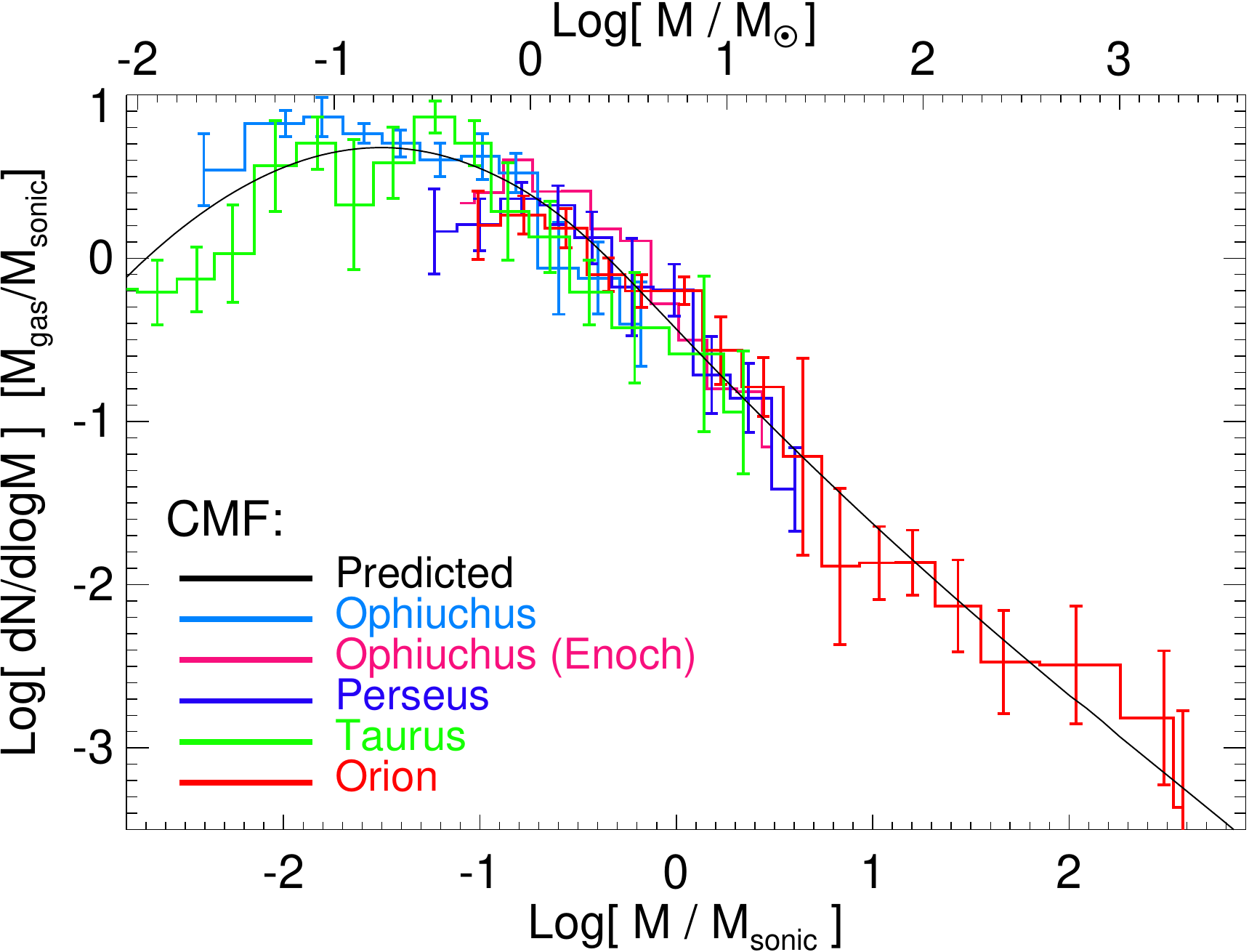}{0.95}
    \caption{Predicted last-crossing mass function from Fig.~\ref{fig:imf}, 
    compared to observed pre-stellar/starless core MFs in different regions. 
    The CMFs for Ophiuchus, Perseus, Taurus, and Orion are compiled in \citet{sadavoy:2010.cmfs}; we also compare the independent determination for Ophiuchus in \citet{enoch:2008.core.mf.clustering}. The CMFs in the Pipe \citep{rathborne:2009.cmf} and Chameleon I \citep{belloche:2011.chameleon.cmf} also agree well but would be indistinguishable from the other systems on this plot. The normalization of the observed/predicted CMF is arbitrary (depending on absolute mass and volume) but -- unlike the comparison with the stellar IMF -- the absolute mass value is not rescaled here. Note the dynamic range is limited, and the observations have substantial uncertainties, but the agreement is good.
    \label{fig:cmf}}
\end{figure}

\vspace{-0.5cm}
\section{The Last-DownCrossing Distribution}
\label{sec:lastcross}

Our derivation here closely follows that of the 
first-crossing distribution in \citet{zhang:2006.general.moving.barrier.solution}, to whom we refer for more details.

Consider the Gaussian field $\delta({\bf x}\,|\,R)$ (which 
for this problem represents the logarithmic density field smoothed in a kernel of 
radius $R$ about ${\bf x}$); the variance in the field $\equiv S(R)$
is a monotonically decreasing function of $R$ so we can take 
$S$ as the independent variable and consider $\delta({\bf x}\,|\,S)$. 
For convenience we will drop the explicit notation of ${\bf x}$. 
The PDF of $\delta(S)$ is just 
\be
P_{0}(\delta\,|\,S) = \frac{1}{\sqrt{2\pi\,S}}\,\exp{\left(-\frac{\delta^{2}}{2\,S} \right)}
\ee

The barrier $B(S)$ is the minimum value $\delta(S)$ which defines 
objects of interest (e.g.\ collapsing regions). 
Normally, we would define the first-crossing distribution by 
beginning with the initial condition $\delta(0)=0$ at $S=0$ ($R\rightarrow\infty$), 
then evaluating $\delta(S)$ at successively increasing $S$ (smaller scales $R$).
This is straightforward: 
given an ``initial'' value $\delta_{0}(S_{0})$ at scale $S_{0}$, the 
probability of a value $\delta_{1}(S_{1})$ at a scale $S_{1}>S_{0}$ is just 
\be
P_{10}[\delta_{1}(S_{1})\,|\,\delta_{0}(S_{0})] = P_{0}(\delta_{1}-\delta_{0}\,|\,S_{1}-S_{0})\ .
\ee
This can be integrated until the ``trajectory'' (or random walk) $\delta(S)$ first exceeds $B(S)$. 

For the last crossing distribution, we need to determine the smallest scale 
at which a trajectory that had previously crossed the barrier at some larger scale 
again falls {below} $B(S)$. 
Consider a trajectory, therefore, in the opposite direction: beginning at 
an initial $\delta_{i}(S_{i})$ at some 
arbitrarily small $R_{i}\rightarrow0$, with a corresponding $S_{i}(R_{i})$, 
and evaluating $\delta$ at successively {\em smaller} $S$ ({\em larger} $R$).
For the last-crossing distribution to be meaningfully defined, it must be the case 
that as $R\rightarrow0$, the probability of $\delta_{i}$ exceeding 
$B(S_{i})$ vanishes, $P_{0}(B[S_{i}]\,|\,S_{i})\rightarrow0$. 

Define the last crossing distribution $\flast(S\,|\,\delta_{i})\,{\rm d}S$ as the probability that a 
trajectory, beginning from this state, crosses $B(S)$ for the first time 
between $S$ and $S+{\rm d}S$. Also define 
$\Pi(\delta\,|\,\delta_{i},\,S)\,{\rm d}\delta$ as 
the probability for the trajectory beginning at $\delta_{i}(S_{i})$ to have a value between 
$\delta$ and $\delta+{\rm d}\delta$ at scale $S$, {\em without} having crossed $B(S)$ at 
any larger $S$. 
Clearly, the integral of $\flast$ 
and integral of $\Pi$ for $\delta<B(S)$ 
must sum to unity: 
\be
\label{eqn:integral.eqn}
1 = \int_{S}^{S_{i}}{\rm d}\Sprime\,\flast(\Sprime\,|\,\delta_{i}) + 
\int_{-\infty}^{B(S)}\Pi(\delta\,|\,\delta_{i},\,S)\,{\rm d}\delta
\ee

For a step in this ``opposite'' direction of decreasing $S$, the 
probability of going from an initial $\delta_{1}(S_{1})$ 
to a value $\delta_{0}(S_{0})$ with $S_{1}>S_{0}$ is related to the probability 
of going from $\delta_{0}(S_{0})$ to $\delta_{1}(S_{1})$ by 
Bayes's theorem: 
\begin{align}
\nonumber P_{01}[\delta_{0}(S_{0})\,|\,\delta_{1}(S_{1})] &= 
P_{10}[\delta_{1}(S_{1})\,|\,\delta_{0}(S_{0})]\,
\frac{P_{0}(\delta_{0}\,|\,S_{0})}{P_{0}(\delta_{1}\,|\,S_{1})}\\
= &P_{0}[\delta_{0}-\delta_{1}\,(S_{0}/S_{1})\,|\,(S_{1}-S_{0})\,(S_{0}/S_{1})]
\end{align}

If we ignored the barrier, $\Pi(\delta\,|\,\delta_{i},\,S)$ 
would be equal to $P_{01}[\delta(S)\,|\,\delta_{i}(S_{i})]$. 
But we must subtract from this the probability that a trajectory 
crosses the barrier at some larger $\Sprime>S$ and then 
passes through $\delta(S)$, so 
\begin{align}
\nonumber \Pi(\delta\,|\,\delta_{i},\,S) &= 
P_{01}[\delta(S)\,|\,\delta_{i}(S_{i})] -  \\
&\int_{S}^{S_{i}}{\rm d}\Sprime\,\flast(\Sprime\,|\,\delta_{i})\,
P_{01}[\delta(S)\,|\,\delta^{\prime}(\Sprime)=B(\Sprime)]
\end{align}

Before going further, note that we do not know the value of 
$\delta_{i}(S_{i})$, but know its distribution. 
The fraction of trajectories crossing the barrier at each $S$ is the 
integral of $\flast(S\,|\,\delta_{i})$ weighted by the distribution 
of $\delta_{i}$. 
We therefore define
\be
\flast(S) \equiv \langle f(S\,|\,\delta_{i})\rangle = 
\int_{-\infty}^{\infty}{\rm d}\delta_{i}\,P_{0}(\delta_{i}\,|\,S_{i})\,\flast(S\,|\,\delta_{i})
\ee
Technically, the upper limit of this integral should be 
$B(S_{i})$, since the trajectory must begin as uncollapsed, but 
since we choose $S_{i}$ such that the collapsed fraction is vanishingly 
small, we can safely take $B(S_{i})\rightarrow\infty$. 
If we take the integral $\int_{-\infty}^{\infty}{\rm d}\delta_{i}\,P_{0}(\delta_{i}\,|\,S_{i})$ 
with respect to both sides of Equation~\ref{eqn:integral.eqn}, 
note that both $\delta$ and $\Sprime$ are independent of $\delta_{i}$, 
and use the fact that 
\be
\int_{-\infty}^{\infty}{\rm d}\delta_{i}\,P_{0}(\delta_{i}\,|\,S_{i})\,P_{01}[\delta(S)\,|\,\delta_{i}(S_{i})]
=P_{0}(\delta\,|\,S)
\ee
we obtain the $\delta_{i}$-averaged equations 
\begin{align}
\label{eqn:int.2}
1 &= \int_{S}^{S_{i}}{\rm d}\Sprime\,\flast(\Sprime) + 
\int_{-\infty}^{B(S)}\Pi(\delta\,|\,S)\,{\rm d}\delta \\ 
\label{eqn:Pi.2}
\Pi(\delta\,|\,S) &= 
P_{0}(\delta\,|\,S) - \int_{S}^{S_{i}}{\rm d}\Sprime\,\flast(\Sprime)\,
P_{01}[\delta(S)\,|\,B(\Sprime)]
\end{align}

Taking the derivative of Eq.~\ref{eqn:int.2}, we obtain
\be
\label{eqn:flast.1}
\flast(S) = \Pi(B(S)\,|\,S)\,\frac{{\rm d}B}{{\rm d}S} + 
\int_{-\infty}^{B(S)}\frac{\partial \Pi(\delta\,|\,S)}{\partial S}\,{\rm d}\delta
\ee

Finally, we insert Eq.~\ref{eqn:Pi.2} in Eq.~\ref{eqn:flast.1},
perform some simplifying algebra\footnote{Specifically, using the relations 
\begin{align}
&\left[\int_{-\infty}^{B(S)}{\rm d}\delta\,P_{01}[\delta(S)\,|\,B(\Sprime)]
\right]{\Bigr|}_{\Sprime\rightarrow S} = \frac{1}{2}\\
&\int_{-\infty}^{B(S)}\frac{\partial P_{0}(\delta\,|\,S)}{\partial S}\,{\rm d}\delta = 
-\frac{B(S)}{2\,S}\,P_{0}(B(S)\,|\,S) \\ 
\nonumber &\int_{-\infty}^{B(S)}\frac{\partial P_{01}[\delta(S)\,|\,B(\Sprime)]}{\partial S}\,{\rm d}\delta = \\
&\ \ \ \ \ \ \ \ \ \ \ -\left[\frac{B(\Sprime)-B(S)}{2\,(\Sprime-S)}+\frac{B(S)}{2\,S}\right]\,P_{01}[B(S)\,|\,B(\Sprime)]
\end{align}
}
and obtain
\begin{align}
\label{eqn:flast}
\flast(S) = g_{1}(S) + \int_{S}^{S_{i}}\,{\rm d}S^{\prime}\,\flast(S^{\prime})\,g_{2}(S,\,S^{\prime})
\end{align}
where 
\begin{align}
g_{1}(S) &= {\Bigl [}2\,\frac{dB}{dS} -\frac{B(S)}{S}{\Bigr]}\,P_{0}(B(S)\,|\,S)\\
g_{2}(S,\,\Sprime) &= {\Bigl[}\frac{B(S)-B(\Sprime)}{S-\Sprime} 
+\frac{B(S)}{S}-2\,\frac{dB}{dS}{\Bigr]}\,P_{01}[B(S)\,|\,B(\Sprime)]
\end{align}

Equation~\ref{eqn:flast} is a Volterra integral equation, 
which for a general barrier $B(S)$ has a unique solution that can be 
calculated by standard numerical methods. For example, 
if we grid $S$ on a mesh with equal spacing, 
\be
S_{n} = S_{i} - n\,\Delta S,\ \ \ n=0,\,1,\,...,\,N,\ \ \ \Delta S= \frac{S_{i}-S}{N}
\ee
and treat $\flast(S_{n})$ as a vector, 
then the integral equation becomes a triangular matrix equation which can 
be solved iteratively: 
\begin{align}
\flast(S_{0}) &= g_{1}(S_{0}) = 0 \\ 
\flast(S_{1}) &= g_{1}(S_{1})\,(1 - H_{1,1})^{-1} \\ 
\flast(S_{n})|_{n>1} &= \frac{g_{1}(S_{n}) + \flast(S_{n-1})\,H_{n,n} + 
\sum\limits_{m=1}^{n-1}\,[\flast(S_{m})+\flast(S_{m-1})]\,H_{n,m}}{1 - H_{n,n}}
\end{align}
where 
\be
H_{n,m} = \frac{\Delta S}{2}\ g_{2}{\Bigl[}S_{n},\,S_{m} + \frac{\Delta S}{2} {\Bigr]}
\ee

For a linear barrier, $B(S)=B_{0}+\beta\,S$, this has a closed-form solution: 
\be
\flast(S\,|\,B=B_{0}+\beta\,S) = \beta\,P_{0}(B(S)\,|\,S) = 
\frac{\beta}{\sqrt{2\pi S}}\exp{\left(-\frac{B^{2}}{2\,S}\right)}
\ee

Equation~\ref{eqn:flast} is qualitatively similar to the governing equation 
for the first-crossing distribution (compare Eq.~5 in \citealt{zhang:2006.general.moving.barrier.solution}), but with some critical differences. Up to a sign, $g_{1}(S)$ is identical.
In the $g_{2}$ term, however, there is an additional 
$B(S)/S$ in the coefficient, and $P_{01}[B(S)\,|\,B(\Sprime)]$ 
appears instead of $P_{10}[B(S)\,|\,B(\Sprime)]$. And of course, the integration 
proceeds in the opposite direction. These corrections 
result in qualitatively different behavior. For example, for the linear barrier, the first-crossing 
distribution has a pre-factor $B_{0}/S$ instead of $\beta$; for a constant barrier ($\beta=0$), 
the first-crossing distribution is well-defined but 
the last-crossing distribution vanishes because there are continued crossings on all 
scales as $S\rightarrow S_{i}$.

\vspace{-0.5cm}
\section{The Core Mass Function: Rigorous Solutions}
\label{sec:solutions}

We have now derived the rigorous solution for the number of bound objects 
per interval in mass $M$, defined as the mass on the {\em smallest} scale on 
which they are self-gravitating. To apply this to a physical system, we need 
the collapse barrier $B(S)$ and variance $S=\sigma^{2}(R)=\sigma^{2}(M)$. 
In \HC, $B(S)$ is defined by the Jeans over-density 
$\rho_{\rm crit}(R) > (c_{s}^{2}+v_{t}^{2}(R))/4\pi\,G$, but the normalization of the
background density, $c_{s}$, and $v_{t}$ is essentially arbitrary. 
Moreover, $S$ is not derived, but a simple phenomenological model is used, 
and the authors avoid uncertainties related to this by dropping terms 
with a derivative in $S$. 
In \paperone, we show how $S(R)$ and $B(S)$ 
can be derived self-consistently on all scales for a galactic 
disk. For a given turbulent power spectrum, together with the assumption 
that the disk is marginally stable ($Q=1$), $S(R)$ can be calculated 
by integrating the contribution from the velocity variance on all scales $R^{\prime}>R$:
\begin{align}
S(R) &= \int_{0}^{\infty} \sigma_{k}^{2}(\mathcal{M}[k])\,|W(k,\,R)|^{2}\,{\rm d}\ln{k} \\ 
\sigma_{k}^{2} &= \ln{{\Bigl [}1 + \frac{3}{4}\,
\frac{v_{t}^{2}(k)}{c_{s}^{2} + \kappa^{2}\,k^{-2}} 
{\Bigr]}} 
\end{align}
where $W$ is the window function for the density smoothing (for convenience, 
we take this to be a $k$-space tophat inside $k<1/R$). 
This is motivated by and closely related to the correlation between Mach number and dispersion in 
turbulent box simulations \citep[see][]{padoan:1997.density.pdf,passot:1998.density.pdf,federrath:2008.density.pdf.vs.forcingtype,price:2011.density.mach.vs.forcing}. 
$B(R)$ is properly given by 
\be
B(R) = \ln{\left(\frac{\rho_{\rm crit}}{\rho_{0}} \right)} + \frac{S(R)}{2}
\ee
\begin{align}
\label{eqn:rhocrit}
\frac{\rho_{\rm crit}}{\rho_{0}} \equiv \frac{Q}{2\,\tilde{\kappa}}\,\left(1+\frac{h}{R} \right)
{\Bigl[} \frac{\sigma_{g}^{2}(R)}{\sigma_{g}^{2}(h)}\,\frac{h}{R}  + 
\tilde{\kappa}^{2}\,\frac{R}{h}{\Bigr]} 
\end{align}
where $\rho_{0}$ is the mean midplane density of the disk, 
$\tilde{\kappa}=\kappa/\Omega=\sqrt{2}$ for a constant-$V_{c}$ disk, 
and 
\be
\sigma_{g}^{2}(R) = c_{s}^{2} + v_{\rm A}^{2} + \langle v_{t}^{2}(R) \rangle 
\ee 
The mapping between radius and mass is 
\be
M(R) \equiv 4\,\pi\,\rho_{\rm crit}\,h^{3}\,
{\Bigl[}\frac{R^{2}}{2\,h^{2}} + {\Bigl(}1+\frac{R}{h}{\Bigr)}\,\exp{{\Bigl(}-\frac{R}{h}{\Bigr)}}-1 {\Bigr]}
\ee
It is easy to see that on small scales, these scalings reduce to the Jeans criterion 
for a combination of thermal ($c_{s}$), turbulent ($v_{t}$), and magnetic ($v_{\rm A}$) 
support, with $M=(4\pi/3)\,\rho_{\rm crit}\,R^{3}$; on large scales it becomes the Toomre criterion 
with $M=\pi\Sigma_{\rm crit}\,R^{2}$.

Finally, we note that because the trajectories $\delta({\bf x}\,|\,R)$ 
defined above sample the Eulerian volume, 
the mass function is given by 
\be
\frac{{\rm d}n}{{\rm d}M} = 
\frac{\rho_{\rm crit}(M)}{M}\,\flast(M)\,{\Bigl |}\frac{{\rm d}S}{{\rm d}M} {\Bigr |}
\ee

It is worth noting that, with $S(R)$ and $B(R)$ derived above, 
there are only two free parameters that together completely specify the model 
in dimensionless units. These are the spectral index $p$ of the turbulent velocity 
spectrum, $E(k)\propto k^{-p}$ (usually in the narrow range $p\approx5/3-2$), 
and its normalization, which we parameterize as the Mach number on large scales 
$\mathcal{M}_{h}^{2}\equiv \langle v_{t}^{2}(h)\rangle/(c_{s}^{2}+v_{\rm A}^{2})$. 
Of course, we must specify the dimensional parameters $h$ (or $c_{s}$) and $\rho_{0}$ 
to give absolute units to the problem, but these simply rescale the predicted quantities.

\begin{figure}
    \centering
    \plotonesize{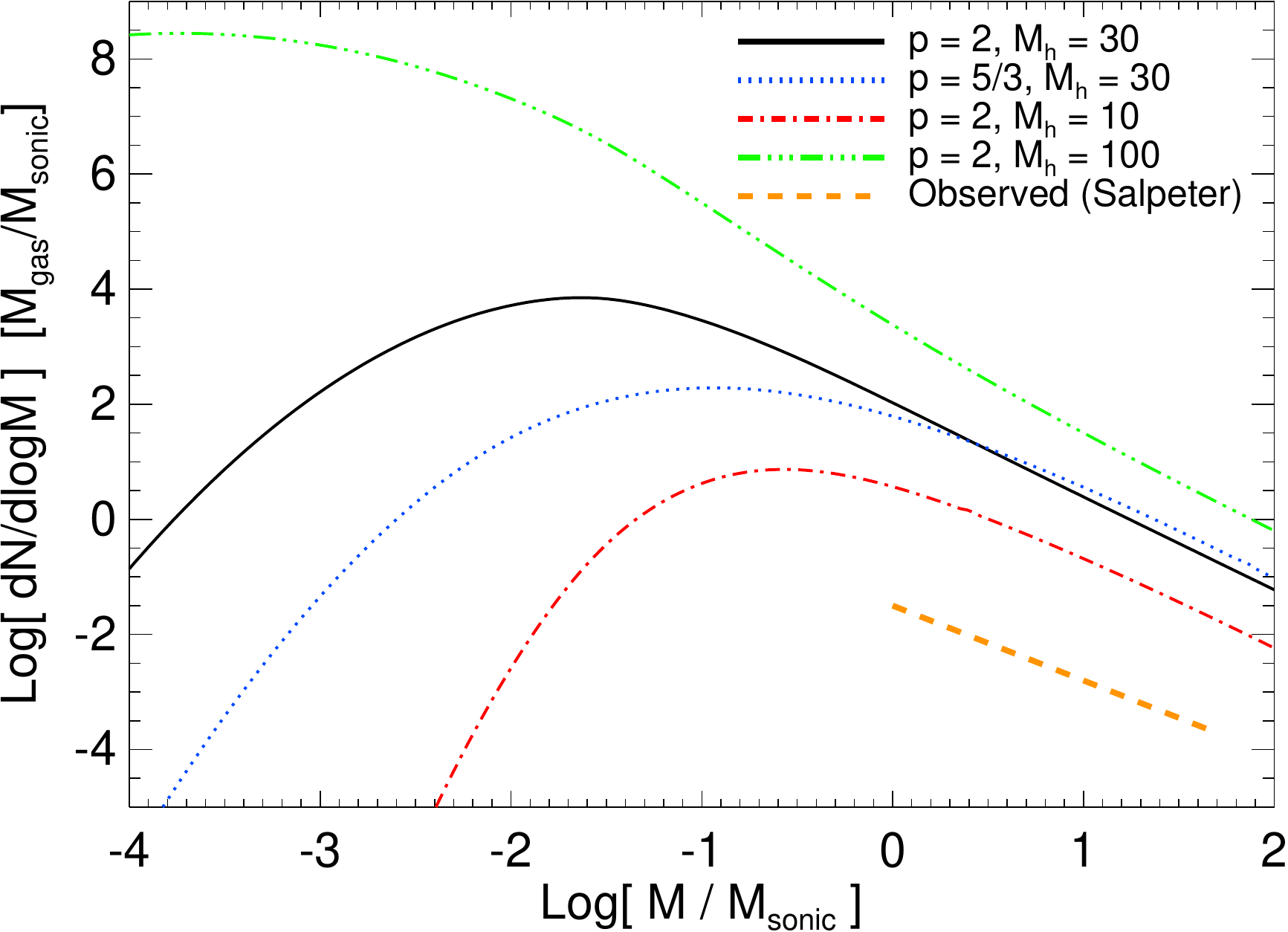}{0.95}
    \caption{Variation in the predicted CMF/last-crossing MF 
    with model assumptions. 
    We compare the standard model from Fig.~\ref{fig:imf} ($p=2$, $\mathcal{M}_{h}=30$). 
    Raising/lowering $\mathcal{M}_{h}$ substantially slows/steepens the cutoff below 
    the sonic mass. Assuming $p=5/3$ (at fixed $\mathcal{M}_{h}$) also slows 
    this cutoff. These variations have almost no effect on the high-mass slope, which is 
    close to Salpeter in all cases. 
    \label{fig:imf.var}}
\end{figure}

\vspace{-0.5cm}
\section{Results}
\label{sec:results}

Using these relations, we are now in a position to calculate the last-crossing 
mass function. 
Figure~\ref{fig:imf}  
shows the results of calculating the last-crossing distribution $\flast(S)$ 
for typical parameters $p=2$ (Burgers' turbulence, typical of highly super-sonic 
turbulence) and $\mathcal{M}_{h}=30$ (typical for GMCs). 
First, we can confirm our analytic derivation in Eq.~\ref{eqn:flast}.
It is easy to calculate the last-crossing distribution by generating a Monte Carlo 
ensemble of trajectories $\delta(R)$ in the standard manner of the 
excursion set formalism (beginning at $R\rightarrow\infty$ and $S=0$), 
and the details of this procedure are given in \paperone; here we simply 
record the last down-crossing for each trajectory and construct the mass function. 
The result is statistically identical to the exact solution. However, below the ``turnover'' 
in the last-crossing mass function, the Monte Carlo method becomes 
extremely expensive and quite noisy, because an extremely small fraction of the 
total galaxy {\em volume} is in low-mass protostellar cores 
(sampling $\ll0.1\,M_{\rm sonic}$ requires $\sim10^{10}$ trajectories).

We compare the last-crossing mass function to the predicted first-crossing distribution -- i.e.\ the mass function of 
bound objects defined on the {\em largest} scale on which they are self-gravitating.
The two are strikingly different: they have different shapes, different power-law slopes, 
and the ``characteristic masses'' are separated by more than six orders of magnitude!
Clearly, it is critical to rigorously address the ``cloud in cloud'' problem 
when attempting to define either. 

Physically, in \paperone\ we argue that the first-crossing distribution should be 
associated with the mass function of GMCs, and show that it agrees very well 
with observations of the same.
The last-crossing distribution, on the other hand, should 
correspond to the proto-stellar core MF -- and by extension to the stellar IMF (although that conversion is uncertain and may well be mass-dependent, as discussed below). We see that here. 
As shown with the approximate derivation in \HC, 
the exact predicted mass function reproduces all of the key 
observed features of the core/stellar MF. The high-mass behavior is an 
approximate power-law with a Salpeter-like slope ${\rm d}N/{\rm d}M\propto M^{-\alpha}$ 
with $\alpha$ slightly larger than $2$; at lower masses the slope flattens and there 
is a subsequent low-mass turnover in good agreement with a 
\citet{kroupa:imf} or \citet{chabrier:imf} IMF. In Figure~\ref{fig:cmf}, we plot a direct comparison between the predicted CMF and observed starless/pre-stellar CMFs in different regions, which are less well-constrained than the stellar IMF but should correspond more closely with what we actually predict. The same behaviors are observed and agree well with our prediction without the need to invoke an uncertain rescaling factor \citep[see e.g.][]{stanke:2006.core.mf.clustering,enoch:2008.core.mf.clustering,rathborne:2009.cmf,sadavoy:2010.cmfs,belloche:2011.chameleon.cmf}.

Why do \HC\ recover similar behavior, given that 
we have shown the first-crossing and last-crossing distributions can 
be so different? The generating equation for the 
mass function in \HC\ is the same as 
that of the original \citet{pressschechter} derivation of the halo 
MF; it neglects the cloud-in-cloud problem (treating it as an {ad hoc} re-normalization). 
Modulo that renormalization, this is identical to the $g_{1}$ term in 
Eq.~\ref{eqn:flast}. This is a good approximation when 
the integral/$g_{2}$ term in Eq.~\ref{eqn:flast} is small and/or proportional to $\flast(S)$. 
This occurs when ${\rm d}B/{\rm d}S\gg B(S)/S\gg 1$; 
if the barrier decreases more rapidly than $S$ (as $R$ increases), then $\delta(R)$ 
for a given trajectory will evolve slowly 
relative to $B(R)$ and the probability of multiple crossings becomes small. 
In Eq.~\ref{eqn:flast}, the $P_{01}$ term approaches a Dirac $\delta$-function 
and the integral term $\rightarrow -\flast(S) + \mathcal{O}({\rm d}B/{\rm d}S)^{-1}$, 
giving $\flast(S) \approx (1/2)\,g_{1}(S) + \mathcal{O}({\rm d}B/{\rm d}S)^{-1}$. 
At scales $R/h\ll1$, relevant for the stellar IMF, $S(R)$ is a weak function 
of $R$ since most of the contributions to $S$ come from the larger Mach 
numbers on larger scales, so this is satisfied.\footnote{This is very different 
from the reason that the original \citet{pressschechter} mass function is correct 
to within a constant normalization, and from the nominal justification 
given in \HC\ for assuming the same. That argument relies on a {\em constant} 
barrier (${\rm d}B/{\rm d}S\equiv0$) 
and applies only to the {\em first}-crossing distribution, for which the symmetry 
in $P_{10}$ (broken in $P_{01}$) 
means that a trajectory at $\delta(S)=B(S)=B_{0}$ is equally likely to 
rise above or fall below the barrier.}

This approximation 
would break down severely at larger scales -- in fact, the \HC\ mass function, 
applied directly to the entire galaxy disk, would include a large part of the 
GMC mass function, and give the result that most of the {\em mass} 
in the core MF is in $\sim10^{6}\,\msun$ ``objects.'' But since the derivation in \HC\ 
specifically (and sensibly) truncates the problem before going to any large scales (focusing only on 
the regime $R/h\ll1$ and $R/R_{\rm GMC}\ll1$), this is avoided. Also, the 
normalization is different, but that is essentially treated as arbitrary by the authors.

So over a limited regime, we can assume 
${\rm d}B/{\rm d}S\gg B(S)/S\gg 1$, ignore the running of $S(R)$, 
and approximate the mass function as 
\be
\label{eqn:approx}
\frac{{\rm d}n}{{\rm d}M}
\sim \frac{\rho_{\rm crit}(M)}{M^{2}\sqrt{2\pi S_{0}}}
{{\Bigl |}\frac{{\rm d}\ln{\rho_{\rm crit}}}{{\rm d}\ln{M}}{\Bigr |}}\,
\exp{\left[-\frac{(\ln{[\rho_{\rm crit}/\rho_{0}]}+S_{0}/2)^{2}}{2\,S_{0}} \right]}
\ee

In the high-mass regime, the power-law slope is set by the 
run of $\rho_{\rm crit}$ with $R$ in the turbulence-dominated regime; 
for $R$ in the range such that $v_{t}\gg c_{s}$ and $R\ll h$, 
$S(M)\sim$\,constant and $\rho_{\rm crit}\propto R^{p-3}$, $M\propto R^{p}$, so 
the approximate analytic scaling in Eq.~\ref{eqn:approx} gives a slope 
\be
\alpha_{\rm highmass} \approx \frac{3\,(1+p^{-1})}{2} 
+ \frac{(3-p)^{2}\,\ln{(M/M_{0})}-p\,\ln{2}}{2\,S(M)\,p^{2}} 
\ee
where $M_{0}\sim \rho_{0}\,h^{3}\sim 10^{5}-10^{6}\,\msun$ for typical conditions.
Based on the derivation in \HC, we would expect a somewhat ``lognormal-like'' 
turnover in the high-mass end of the CMF; however, this does not self-consistently 
calculate the running of $S(M)$. In the exact calculation, the correction term $\propto 1/S(M)$ 
is already small in the range of interest ($\approx -0.1$, for $S(M)\sim10$ and $M/M_{0}\sim10^{-6}-10^{-4}$) and higher-order corrections nearly cancel, giving a very nearly Salpeter ($\alpha\approx2.2-2.4$) power-law behavior over $\sim2\,$dex in mass.

The turnover occurs below the sonic length, where thermal support both makes 
the equation of state ``stiffer'' and suppresses the running of the variance 
$S(R)$, giving $\rho_{\rm crit}\propto R^{-2}$, $M\propto R$, so 
\be
\alpha_{\rm lowmass}\approx 3 -\frac{1.5+4\ln{\mathcal{M}_{h}}-2\ln{(M/M_{0})}}{S(M)}
\ee
For typical parameters this becomes $\alpha_{\rm lowmass}\sim-2-0.16\,\ln{(M/M_{\rm sonic})}$. 

Our solution also allows us to predict the normalization and mass scale of the 
mass function, without having to make any {\em ad hoc} assumptions 
to renormalize the problem about a given scale. The CMF begins to turn 
over below the sonic length, $R_{\rm sonic}=h\,\mathcal{M}_{h}^{-2/(p-1)}$, 
with mass $M_{\rm sonic}\approx (4\sqrt{2}\pi/3)\,\mathcal{M}_{h}^{-2p/(p-1)} M_{0}
= (2/3)\,c_{s}^{2}\,R_{\rm sonic}/G \approx \msun\,(c_{s}/0.3\,{\rm km\,s^{-1}})^{2}\,
(R_{\rm sonic}/0.1\,{\rm pc})$.
The mass scale of the CMF the stellar IMF is therefore a natural consequence of Jeans collapse in a turbulent medium, with characteristic masses that depend only on the sound speed and sonic length (themselves related). 
This is a well-established result from both analytic work and numerical simulations \citep[see e.g.][]{klessen:2000.cluster.formation,klessen:2001.sf.cloud.pdf,bate:2005.imf.mass.jeansmass}, and is true even if the gas is not isothermal \citep[although our sub-sonic extrapolation is questionable in this case; see][]{larson:1985.imf.scale,larson:2005.imf.scale.thermalphysics,jappsen:2005.imf.scale.thermalphysics}.  There is, of course, some efficiency factor $\epsilon_{\rm core}$ that relates the mass of a collapsing, bound protostellar ``core'' (which is what we actually predict) 
to the mass of a star. This may well depend on mass (and the thermal physics below the sonic length), which will introduce additional corrections between the IMF here and the stellar IMF (see \S~\ref{sec:cmf.imf} below).
However, the calculation here suggests that such corrections should 
be order unity, and arguments from 
observations and models of outflows similarly suggest a modest 
$\epsilon_{\rm core}\sim0.5$ \citep{matzner:2000.lowmass.sf.eff,stanke:2006.core.mf.clustering,alves:2007.core.mf.clustering,enoch:2008.core.mf.clustering}.
If constant (a significant assumption), this is comparable to the normalization uncertainty 
in $M_{\rm sonic}$ from $c_{s}$ and $R_{\rm sonic}$.

In Fig.~\ref{fig:imf.var}, we plot the predicted last-crossing distribution 
as a function of model parameters. We compare our ``default'' model 
($p=2$, $\mathcal{M}_{h}=30$) to one with $p=5/3$. For otherwise equal parameters, 
this shifts $M_{\rm sonic}$ to lower absolute values, but in units of the sonic mass 
the behavior is qualitatively the same. The high-mass slope is nearly identical -- the difference 
predicted by the approximate Eq.~\ref{eqn:approx} or in \HC\ is small to begin 
with, but is also largely canceled out by the second-order corrections from the proper last-crossing 
distribution. The low-mass turnover is much ``slower,'' however, since the 
turbulent velocity declines more slowly relative to the sound speed ($v_{t}\propto R^{1/3}$ 
instead of $R^{1/2}$ for $p=2$), which enters into both the barrier and the run of $S(R)$. 
Raising the Mach number on large scales, $\mathcal{M}_{h}$, has a similar effect, and 
slightly steepens the high-mass slope near the turnover mass.
Decreasing the Mach number steepens the turnover below $M_{\rm sonic}$, and 
for quite low $\mathcal{M}_{h}$ also begins to manifest a high-mass cutoff. 
These are the dominant physical effects. Changing more subtle model choices -- for example, 
using a Gaussian window function to smooth the density field instead of a $k$-space tophat, 
make relatively little difference.

\vspace{-0.5cm}
\section{Caveats on The CMF-IMF Conversion}
\label{sec:cmf.imf}

What we have calculated here is the mass function of self-gravitating, non-fragmenting cores at a given ``snapshot'' or instant, for a galaxy with fully developed non-linear turbulence. It is important to note that this is not yet the stellar IMF. As noted above, it is necessary to invoke some mean efficiency of $\sim30\%$ in the conversion of core mass to stellar mass -- presumably this comes from some combination of outflows and accretion histories, which could easily depend on mass; it is also possible that accreting protostars can be ejected from their environments, introducing a large core-to-stellar mass scatter. None of these physics are included in the current model; it may be possible to add them ``on top'' of the model here, but to do so it would be necessary to construct a more complex analytic model with some significant additional assumptions about how pre-stellar cores contract and grow, and how  outflows affect the ambient medium. 

Although within the ``snapshots'' here, the cores have no sub-scales which are independently self-gravitating, there can certainly be successive fragmentation on smaller scales within the cores and/or stellar accretion disks as the cores contract and form stars \citep[which will depend on the gas thermodynamics and stellar feedback; see references above and e.g.][]{krumholz:2009.massive.sf.accretion,peters:2010.massive.sf.accretion}. Indeed this must occur, as a large fraction of stars are in binaries and high-mass stars are almost always members of higher order multiple stellar systems \citep{lada:2006.low.binary.fraction,zinnecker:2007.high.binary.frac.massivestars}. If the multiplicity is mass-dependent (i.e.\ fragmentation is not scale-free), this will also change the mass spectrum from CMF to IMF. There may be some prospects of including this in future, time-dependent excursion set models (discussed below), but if fragmentation is feedback-dependent or occurs within the stellar accretion disk, following it accurately depends on physics not included in the current model.

Even if each core forms a fixed fraction of its mass into a fixed number of stars, the timescales for it to do so may be different. What the IMF really samples is the relative ``formation rate'' of stars of a given mass; so if the timescale for collapse of massive cores is much longer than the timescale for collapse of low-mass cores, this can significantly change the shape of the IMF relative to the CMF \citep[see the discussion in][]{clark:2007.cmf.imf.timescale.conversion}. At high masses $M\gtrsim M_{\rm sonic}$, we do not expect the dependence of collapse timescale on mass to strongly change our results. For $R\gtrsim R_{\rm sonic}$ and $R\ll h$, the characteristic collapse time for cores scales as $\sim t_{\rm dyn} \propto \rho_{\rm crit}^{-1/2} \propto R/v_{t}  \propto R^{1/2}\propto M^{1/4}$. If the rate of star formation simply scales as $\propto n_{\rm core}/t_{\rm collapse}$, this implies the bright end of the stellar IMF will be steeper than that of the CMF predicted here by $M^{-1/4}$. This is not a negligible correction, but it is also not very large (comparable to the difference between $p=5/3$ and $p=2$). Moreover, for $p=2$ it actually goes in the ``right'' direction: the CMF slope predicted (with second-order corrections, $\alpha\sim2.15-2.2$) is slightly more shallow than the canonical Salpeter slope, so this would bring the predicted IMF slope closer to observations. However, at low masses ($\ll M_{\rm sonic}$), this effect is more important: $t_{\rm dyn} \propto R/c_{s} \propto R \propto M$, so the timescale correction can significantly modify the CMF-IMF relation. But in this regime, we already noted that our predictions are much more sensitive to the turbulent spectrum, and our simplifying assumption that the gas is isothermal, while introducing negligible corrections at $R>R_{\rm sonic}$, can have large effects on CMF in the sub-stellar ($R\ll R_{\rm sonic}$) regime \citep[see][]{hennebelle:2009.imf.variation}. We therefore caution that while collapse times are critical to the CMF-IMF conversion at low masses, the gas thermodynamics is likely to be an equal or greater source of uncertainty (given the simplicity of our derivation here).

\vspace{-0.5cm}
\section{Discussion \&\ Conclusions}
\label{sec:discussion}

We have applied the excursion set model of the ISM, developed in \paperone, 
to predict the mass spectrum of star-forming cores and, by (admittedly less certain) extrapolation, the stellar IMF. 
We note that the density field smoothed on a scale $R$ about a random point ${\bf x}$ 
in the ISM, $\delta({\bf x},\,R)$ may be self-gravitating on many scales. The largest of these scales 
defines the ``first crossing distribution,'' which we show in \paperone\ should be associated 
with the distribution of GMCs. If it is also self-gravitating on smaller scales, then this ``parent'' 
scale is not part of the core MF or stellar IMF, but will fragment into smaller units that collapse first: rather, 
the CMF should be associated with the ``last crossing distribution,'' or the mass function defined 
on the {\em smallest} scales on which systems are self-gravitating. To our knowledge, this has not previously been studied in cosmological or ISM applications. This can be determined 
in a standard Monte Carlo excursion set approach, but we also derive the exact analytic solution 
for an arbitrary collapse threshold. We use this to predict the CMF, and find it agrees very well 
with that observed. The high-mass slope emerges as a consequence of turbulent support/fragmentation and is very close to a power-law over $\sim2$\,dex, and very insensitive to the properties of turbulence for reasonable power spectra. The slope flattens at the sonic length/mass, $\sim c_{s}^{2}\,R_{\rm sonic}/G$, as thermal pressure makes the effective equation of state more ``stiff.'' The shape of this turnover is robust, but exactly how rapidly it occurs (in units of mass) depends on the Mach number and turbulent spectral shape. 

Our calculation supports the conclusions of \HC, who derive the same behavior from a simpler ``collapsed fraction'' argument, and do not attempt to calculate the variance in the density field or disk-scale effects. In many ways, the derivation here is related to that work as that in \citet{bond:1991.eps} is related to \citet{pressschechter}. We have provided a formal mathematical justification, and exact solution, for the approximate mass function therein. Moreover we provide a means to calculate the normalization (of both space density and mass scale) and variance beginning at the parent disk scale, which removes many ambiguities. And incorporating the argument into the full excursion set framework makes possible many additional calculations. 

We stress that the model used here to derive the CMF is {\em exactly} the same as that used 
in \paperone\ to predict the properties of the GMC population and large-scale ISM structure. 
The only difference is that here we examine the last, instead of first, crossing distribution. Thus, beginning at the scale of an entire galactic disk, with the only ``free parameter'' being the turbulent spectrum, we derive first the distribution of GMCs, and {\em simultaneously} the MF of cores on much smaller scales near the sonic length, within those GMCs. This is, to our knowledge, the first analytic model that simultaneously predicts the CMF/IMF and large-scale ISM properties -- both in remarkably good agreement with observations.

There are a number of obvious extensions to our calculation. 
With a rigorous definition of the last-crossing distribution, it is straightforward to 
predict other properties derived for the GMC population in \paperone: 
the core size-mass and linewidth-size relations, the distribution of ``voids'' or ``bubbles,'' 
the collapse rate of bound structures (and rate at which new mass is ``supplied'' 
into cores), and the dependence of these properties on 
turbulent spectral index, Mach number, gas thermal properties, 
and more. With a well-defined 
$\flast$ (the last-crossing distribution; Eq.~\ref{eqn:flast}) is it possible to calculate the correlation function 
of individual stars and clustering properties of star formation.
 
In \paperone, we also show how the excursion set formalism can be used to 
construct a Monte Carlo ``merger/fragmentation tree,'' analogous to extended Press-Schechter 
halo merger trees, with which it is possible to follow the time evolution of 
the ISM and GMC populations and build semi-analytic models for the same.
Since the last-crossing is determined for all 
trajectories in the Monte Carlo formalism, it is straightforward to extend this to sample 
the IMF in a fully rigorous time-dependent manner, as a function of properties 
in different GMCs. With this formalism, GMCs, once bound, can be allowed to depart from 
the background flow, and this can in turn be allowed to change the CMF and ultimately the IMF. The IMF 
calculation could, in turn, be used to determine the per-GMC SFR and 
stellar mass distribution, which then informs whatever model of feedback is used to 
evolve the clouds themselves. One might imagine a model in which, 
for example, the GMC remains bound until a sufficient number of sub-regions have 
experienced last-crossing at a scale sufficiently large to form O-stars, which 
then dissociate the cloud. This also provides a means to generalize calculations 
such as that in \citet{krumholz.schmidt} to rigorously derive the star formation 
{\em rate}, without having to assume an ad hoc collapse timescale $\sim t_{\rm dyn}$. 

It is also possible to follow time-dependent 
fragmentation. One concern with our ``instantaneous'' calculation (which essentially 
calculates last-crossing for the galaxy at a ``snapshot'' in time) is that a non-negligible 
amount of mass has its last crossing at scales much larger than any single star: 
see e.g.\ the overlap in the high-mass $\flast$ and low-mass tail of the first-crossing 
or GMC mass function in Fig.~\ref{fig:imf}.
In a fully time-dependent model though, 
these regions would require at least $\sim1$ dynamical time to contract, 
which amounts to many dynamical times for the denser sub-regions of the cloud. 
In this time, therefore, a large number of sub-regions might cross below 
the barrier -- this is a purely analytic means to calculate time-dependent fragmentation 
of collapsing objects. This is likely to resolve the ``problem'' of these otherwise 
high-mass objects, but also to steepen the high-mass end of the 
predicted IMF relative to the CMF. Extension of our models to a fully time-dependent formulation may also provide a means to address the timescale corrections and fragmentation processes discussed in \S~\ref{sec:cmf.imf}, which are critical to understand the relation between CMF predicted here and stellar IMF. Including the effect of outflows is, however, non-trivial, and will require the introduction of fundamentally new physics in the models.

\vspace{-0.7cm}
\acknowledgments 
We thank Chris McKee and Eliot Quataert for helpful discussions 
during the development of this work. We also thank our referee, Ralf Klessen, as well as Patrick Hennebelle, Gilles Chabrier, and Alyssa Goodman for a number of suggestions and thoughtful comments. Support for PFH was provided by NASA through Einstein Postdoctoral Fellowship Award Number PF1-120083 issued by the Chandra X-ray Observatory Center, which is operated by the Smithsonian Astrophysical Observatory for and on behalf of the NASA under contract NAS8-03060.\\

\vspace{-0.1cm}
\bibliography{/Users/phopkins/Documents/lars_galaxies/papers/ms}

\end{document}